\documentclass[onecolumn,sort&compress,numbers]{els-mrw} 
\usepackage{amsmath,amssymb,amsfonts,amsthm,makeidx,graphicx}
\usepackage{txfonts}
\usepackage{helvet}

\begin{document}

\chapter[Particle Cosmology]{Particle Cosmology}
\label{chap:particlecosmology}

\author[1,2,3]{Venus Keus}
\address[1]{\orgname{Dublin Institute for Advanced Studies}, \orgdiv{School of Theoretical Physics}, \orgaddress{10 Burlington Road, D04 C932, Dublin, Ireland}}
\address[2]{\orgname{Theoretical Physics Department}, \orgdiv{CERN}, \orgaddress{1 Esplanade des Particules, Geneva 23, CH-1211, Switzerland}}
\address[3]{\orgname{Department of Physics and Helsinki Institute of Physics}, \orgdiv{University of Helsinki}, \orgaddress{Gustaf Hallstromin katu 2, FIN-00014, Helsinki, Finland}}

\articletag{Fundamentals of Cosmology and Particle Physics}

\maketitle


\begin{abstract}[Abstract]
Particle cosmology is the branch of science that seeks to understand the birth and evolution of the Universe by applying the principles of particle physics. It brings together the physics of the very small (fundamental particles and forces) with the physics of the very large (the structure and evolution of the cosmos). In many ways, the early Universe acts as a natural laboratory - one far more energetic than any collider we can build - offering unique insights into phenomena that may never be accessible on Earth.
Cosmological observations such as the Cosmic Microwave Background, the distribution of galaxies, and the accelerating expansion of the Universe serve as windows into the fundamental laws of nature. At the same time, theoretical developments in particle physics have led to theories, such as inflation, baryogenesis, and Dark Matter, that help explain key features of the cosmos.
\end{abstract}

\begin{keywords}
Particle cosmology; early Universe; inflation; baryogenesis; Dark Matter; Dark Energy; gravitational waves; phase transitions; Cosmic Microwave Background; beyond the Standard Model.
\end{keywords}

\section*{Objectives}
After reading this chapter, the reader should be able to:

\begin{itemize}
\item Summarise the thermal history of the early Universe and its connection to fundamental physics.

\item Explain how inflation resolves key cosmological puzzles and seeds structure.

\item Understand the Sakharov conditions and leading baryogenesis mechanisms.

\item Compare Dark Matter candidates and their cosmological origins.

\item Describe the evidence for Dark Energy and key theoretical models.

\item Recognise how gravitational waves probe early-Universe phenomena.

\item Appreciate how cosmology and particle physics constrain new physics.
\end{itemize}

\section{Introduction}

Particle cosmology explores how the laws of particle physics shape the evolution of the Universe - from its earliest moments to the cosmic structures we observe today. At its core, it asks some of the most fundamental questions in physics:
\begin{itemize}
\item 
What physical conditions governed the Universe moments after the Big Bang?
\item
How did matter come to dominate over antimatter?
\item
What is the nature of the unseen mass (Dark Matter) and the energy driving cosmic acceleration (Dark Energy)?
\item
What traces of early-universe physics can be detected today - through cosmic radiation, gravitational waves, or relic particles?
\end{itemize}
The answers to these questions lie in the intersection of high-energy particle physics and cosmology. The early Universe, at extreme temperatures and densities, provides the only known setting where fundamental interactions - the strong, weak, electromagnetic, and gravitational forces - operated at their full quantum complexity. It therefore provides a unique and natural laboratory for testing the Standard Model (SM) and probing beyond Standard Model (BSM) frameworks, at energy scales far exceeding those accessible to terrestrial experiments.
This chapter provides a guided introduction to the major themes of particle cosmology. It assumes familiarity with the broad ideas of modern physics but avoids technical detail in favour of conceptual clarity and orientation within the field. Its goal is to serve as a bridge between the specialised literature and researchers from diverse backgrounds - in hadron physics, astrophysics, nuclear and accelerator physics, or even medical physics - who wish to engage with cosmology at a serious level.

\section{Thermal history of the Universe}

The early Universe was governed by extreme conditions, very high temperatures and densities that far exceed those produced in any current experiment. During its first few seconds, the entire cosmos behaved like a high-energy particle collider, in thermal equilibrium, with all species of particles constantly interacting, transforming, and annihilating.
The thermal history of the Universe is a chronological account of how these physical conditions evolved as the Universe expanded and cooled. Understanding this history allows us to trace back how matter formed, how forces separated, and how the structure of the cosmos emerged from an initial state that was nearly featureless.
This section provides a conceptually rich overview of the thermal epochs, phase transitions, and freeze-out processes that shaped the Universe we observe today. It also highlights how these stages connect directly to ideas in particle physics, making this history not only cosmologically informative but also experimentally relevant.

\subsection{Chronology of thermal milestones}

One of the most powerful insights in cosmology is that the expansion of the Universe acts as a ``cosmic thermostat''. As space expands, energy dilutes, and temperature drops. Because the rate of expansion is tied to the energy content of the Universe, the cooling process is not linear but depends on the dominant form of energy at any given time - whether radiation, matter, or vacuum energy.
Early on, the energy density was dominated by relativistic particles, and hence the temperature dropped rapidly. But it did so in a predictable manner, allowing physicists to ``rewind the clock'' and identify key moments in cosmic time by their associated temperatures.
This thermal evolution sets the stage for understanding when different particle processes occurred, when interactions froze out, and when the fundamental forces began to behave distinctly.
Here is a time-line of key thermal events in the early Universe, proceeding from the highest energies to the more familiar physics of the later cosmos. Each epoch offers insight into both cosmology and particle physics~\cite{Kolb:1988aj}:

\subparagraph{The Planck epoch ($<$ $10^{-43}$ seconds)} 
At the earliest conceivable moment, densities were so high that quantum effects of gravity were significant. Unfortunately, we lack a complete theory of quantum gravity, so this epoch remains speculative. However, it marks the natural cut-off for where our current physical theories begin to apply.

\subparagraph{Grand Unification epoch ($\sim 10^{-36}$ to $10^{-32}$ seconds)} 
Many BSM scenarios propose that the strong, weak, and electromagnetic forces were once unified into a single force at ultra-high energies. In Grand Unified Theories (GUTs), this unification is associated with a symmetry that spontaneously breaks as the Universe cools, potentially generating heavy GUT-scale particles and topological defects like magnetic monopoles.

\subparagraph{Inflation and reheating ($\sim 10^{-36}$ to $10^{-32}$ seconds and beyond)} 
Inflation is a period of accelerated expansion during which a small, causally connected patch inflated to become the entire observable Universe, which is naturally flat, homogeneous and isotropic, seeding the very primordial fluctuations that grew into galaxies and cosmic structures
While inflation erases most prior features, it ends in a process called \textit{reheating}, where the energy of the inflaton field is transferred into a hot bath of SM particles. This process resets the thermal clock, and defines the starting point for post-inflationary particle cosmology.

\subparagraph{Electroweak epoch ($\sim 10^{-12}$ seconds; $\sim$ 100 GeV)} 
As the Universe cooled to energies around the electroweak scale, the electromagnetic and weak nuclear forces separated. Before this transition, they behaved as a single force; afterwards, the $W$ and $Z$ bosons acquired mass and the Higgs field acquired its vacuum expectation value.
This epoch may have featured a first-order phase transition, a rapid change in the Higgs vacuum state, accompanied by bubbles of broken symmetry nucleating through the symmetric phase, further expanding to fill the entire space.

\subparagraph{Quark-hadron transition ($\sim 10^{-6}$ seconds; $\sim$ 150 MeV)} 
At this temperature, quarks and gluons which were previously free in a quark-gluon plasma, became confined into hadrons (such as protons and neutrons) through the process of confinement in Quantum Chromodynamics (QCD).
The dynamics of this epoch are intimately tied to the properties of strongly interacting matter.

\subparagraph{Neutrino decoupling and electron-positron annihilation ($\sim 1$ second; $\sim 1$ MeV)} Neutrinos, which were in equilibrium with other species via weak interactions, decoupled from the thermal bath when the expansion rate overtook the interaction rate. From that moment on, they streamed freely through space.
Shortly afterwards, electrons and positrons annihilated into photons. This released energy into the photon bath, causing the photon temperature to exceed that of neutrinos. This effect persists today, and is measurable in the precise ratio of photon-to-neutrino temperatures in the Cosmic Microwave Background (CMB).

\subparagraph{Big Bang Nucleosynthesis ($\sim$ 3 minutes; $\sim$ 0.1 MeV)} 
In this epoch, free neutrons and protons began to combine into light nuclei, such as deuterium, helium-3, helium-4, and lithium-7. The precise abundances of these elements are sensitive to (i) the neutron-to-proton ratio (set by weak interaction rates and freeze-out), (ii) the expansion rate (hence the number of relativistic species, or effective neutrino species), and (iii) any exotic energy injection (e.g., from decaying particles).
Big Bang Nucleosynthesis (BBN) remains one of the most powerful probes of new physics in the MeV energy range, and it provides robust constraints on BSM particles such as sterile neutrinos, axions, or additional relativistic degrees of freedom.

\subparagraph{Recombination and photon decoupling ($\sim$ 380,000 years; $\sim$ 0.3 eV)} 
At this point, temperatures dropped enough for electrons to bind with nuclei to form neutral hydrogen. With fewer free electrons to scatter off of, photons decoupled from matter and began free-streaming through the Universe.
These photons form the CMB, a snapshot of the Universe at this moment of last scattering. The pattern of tiny anisotropies in the CMB provides precise measurements of the contents, geometry, and expansion history of the Universe.

\subsection{Relic abundances and the dynamics of thermal decoupling}
A central concept in the thermal history of the Universe is the effective number of relativistic degrees of freedom, $g_*$, which quantifies how many particle species contribute to the energy density at a given temperature. As the Universe expands and cools, heavier particles become non-relativistic and drop out of thermal equilibrium, leading to a stepwise decrease in $g_{*}$. These transitions affect the expansion rate and leave imprints on key observables, such as the shape and amplitude of the CMB, the timing and outcomes of BBN, and the freeze-out or freeze-in of Dark Matter and other relics. Accurate calculations of $g_*$, often requiring thermal field theory and non-perturbative methods like lattice QCD, form a crucial link between particle physics and cosmological modelling.

A recurring theme in thermal history is the idea of \textit{freeze-out}: a particle species is in equilibrium with the thermal bath, but as the Universe expands and the temperature drops, the interactions that maintain equilibrium become inefficient. The species ``freezes out'' and its abundance becomes fixed - or at least evolves independently of the rest of the plasma.
This process defines the relic abundance of many important particles, for example, neutrinos which decouple at $\sim$1 MeV, or Dark Matter candidates which decouple around the electroweak scale.
Understanding freeze-out mechanisms is essential in connecting particle physics models to cosmological data. It also explains why the early Universe is so powerful as a probe; if a particle was ever in equilibrium and survived freeze-out, its abundance today can be predicted from first principles.

\subsection{Thermal history as a probe of particle physics}
What makes the thermal history of the Universe so compelling is that it is observationally accessible. Although we cannot recreate the extreme conditions of the early cosmos in the laboratory, we can observe its aftermath through the CMB, the abundances of light elements, and, potentially, primordial gravitational waves. We can simulate its microphysics using tools such as lattice QCD, Boltzmann solvers, and effective field theories. And we can use it to constrain or motivate theoretical models in areas ranging from Dark Matter and baryogenesis to neutrino physics. The remarkable level of understanding we already have allows us to pinpoint where new physics is needed, and what kinds of experiments or observations might reveal it. For particle physicists, the thermal history is not merely a backdrop to cosmology; it is a vast testing ground where theories developed for the smallest scales confront the evolution of the Universe.

\section{Inflation and primordial fluctuations}

Modern cosmology rests on the framework of the hot Big Bang theory, which successfully describes the evolution of the Universe from a fraction of a second after its origin to the present day. Yet, despite its empirical successes, this framework leaves several deep questions unresolved. Why is the Universe so homogeneous and isotropic on large scales, even though widely separated regions were never in causal contact? Why is the spatial curvature so close to zero, implying an almost perfectly flat geometry? And why do we observe no exotic relics, such as magnetic monopoles, predicted by certain high-energy theories?
These fine-tuning problems point to the need for a mechanism that dynamically sets the initial conditions of the Universe. The theory of cosmic inflation provides such a mechanism. It proposes a brief period of rapid, accelerated expansion in the first tiny fraction of a second after the Big Bang, during which the size of a small region of space increased exponentially, by at least a factor of $10^{26}$ in volume. This expansion smooths out any initial inhomogeneities, flattens the geometry of space, and stretches causally connected regions to cosmic scales, thereby explaining the observed uniformity of the CMB~\cite{Guth:1980zm,Linde:1981mu}.

Inflation not only resolves these conceptual issues but also seeds the primordial fluctuations that give rise to the large-scale structure of the Universe. Quantum fluctuations of fields during inflation are stretched to macroscopic scales and become frozen as curvature perturbations. These imprints survive as temperature anisotropies in the CMB and the initial conditions for galaxy formation. Thus, inflation serves as a bridge between quantum field theory and cosmological observation, connecting the microphysics of the early Universe to its vast, observable structure.
This section explores the physics and consequences of inflation, the generation of primordial fluctuations, and their observational signatures. It also highlights how inflation offers a powerful framework for testing early-universe particle physics and serves as a gateway into deeper questions at the intersection of cosmology and high-energy theory.

\subsection{Inflationary dynamics and the end of inflation}

While the concept of inflation elegantly resolves several fundamental puzzles in cosmology, its underlying physical origin remains unknown. In most models, inflation is driven by a hypothetical scalar field known as the \textit{inflaton}. The dynamics of this field, particularly the shape of its potential, determine the duration of inflation, the expansion rate of the Universe during this epoch, and the mechanism by which inflation ends. Although the inflaton has not been observed, it may be embedded within a broader theoretical framework such as supersymmetry, grand unification, or string theory. Different inflaton potentials lead to distinct predictions for key observables, including the spectrum of primordial fluctuations, which can be tested through measurements of CMB and large-scale structure.

Inflation must eventually end, and this transition, known as reheating, is crucial for connecting the inflationary phase to the hot Big Bang~\cite{Allahverdi:2010xz}. As the inflaton field decays, its energy is transferred into SM particles, filling the Universe with a hot, dense plasma. This marks the onset of the radiation-dominated era and sets the stage for BBN. The details of reheating, including whether it proceeds via perturbative decay, non-perturbative effects like preheating, or involves non-thermal particle production, can influence the thermal history of the Universe. Reheating also determines how many $e$-folds (the number of times the universe has expanded by a factor of $e$) of inflation are required to match present-day observations, which in turn constrains inflationary models. In some scenarios, reheating may even leave observable relics, such as gravitational waves or exotic particle populations, offering another possible probe into BSM physics.

\subsection{Quantum fluctuations and the origin of structure}
One of the most profound achievements of inflation is its ability to connect quantum mechanics with cosmological structure formation. During inflation, quantum fluctuations in the inflaton field and spacetime curvature are stretched to cosmological scales by the rapid expansion. Once these fluctuations cross the Hubble horizon, they become effectively frozen as classical perturbations in the energy density. After inflation ends, these perturbations seed the primordial inhomogeneities that grow into the large-scale structure of the Universe.
These imprints are observable as temperature anisotropies in the CMB and as the initial conditions for galaxy formation. Thus, inflation not only explains the initial conditions of the hot Big Bang but also provides a framework for understanding the origin of all structure in the Universe, from cosmic filaments to stars, galaxies, and clusters.

This connection between quantum mechanics and cosmology is not merely theoretical. It is tested through precise measurements of the CMB, particularly its temperature fluctuations and polarisation patterns, as well as through the large-scale distribution of galaxies~\cite{Planck:2018jri}. Key observables include the scalar spectral index $n_s$, which measures the scale dependence of the primordial power spectrum. Inflation predicts a nearly scale-invariant spectrum, consistent with current data. Another critical observable is the tensor-to-scalar ratio $r$, which quantifies the relative strength of gravitational waves to density fluctuations and provides a direct handle on the energy scale of inflation. The detection of primordial B-mode polarisation would be a smoking gun for such gravitational waves and remains a major goal of upcoming CMB experiments.
Beyond these leading observables, more subtle features, such as non-Gaussianities, isocurvature perturbations, or features in the power spectrum, can probe more complex models involving multiple fields or non-standard interactions~\cite{Komatsu:2010hc}.

\subsection{Popular inflationary models}

Inflation is not a single theory but a broad class of models, each differing in the form of the inflaton potential, its interactions, and its coupling to gravity~\cite{Martin:2013tda}. Large-field models involve the inflaton traversing super-Planckian distances in field space and typically predict sizeable tensor-to-scalar ratios, making them prime targets for gravitational wave searches. In contrast, small-field models describe inflation near a local maximum of the potential and often lead to smaller gravitational wave signatures. Plateau-like potentials, such as in Starobinsky inflation~\cite{Starobinsky:1982ee}, feature a flat region at large field values and are strongly supported by current data. More complex scenarios include multi-field or hybrid models, which introduce additional degrees of freedom and can produce isocurvature perturbations or richer reheating dynamics.

Inflationary model building sits at the intersection of quantum field theory, general relativity, and high-energy phenomenology. Among minimal constructions, the Higgs inflation model~\cite{Bezrukov:2007ep} suggests that the SM-Higgs field, if non-minimally coupled to gravity, could act as the inflaton. While conceptually economical, such models raise important questions about unitarity and their ultraviolet completion. Other models extend the SM with a small number of scalar fields, aiming to remain as close as possible to known physics. More speculative approaches embed inflation in string theory or frameworks of quantum gravity. Regardless of the underlying theory, inflation provides a unique window into energy scales far beyond the reach of terrestrial accelerators, possibly near the GUT or Planck scale.

\subsection{Future prospects}

Future experiments will subject inflation to increasingly stringent observational tests. Projects like CMB-S4 and LiteBIRD aim to detect primordial B-mode polarisation, a direct signature of gravitational waves from inflation. Large-scale structure surveys, including Euclid and the Rubin Observatory, will map the matter distribution of the Universe with exquisite detail, tightening constraints on primordial fluctuations. Gravitational wave observatories may also detect signals from violent dynamics at the end of inflation, such as preheating or phase transitions. Even in the absence of new detections, improved null results will help rule out broad classes of models and sharpen theoretical frameworks. Inflation remains one of the most elegant and predictive ideas in early-Universe cosmology, offering a rich testing ground for ideas about fields, symmetries, and fundamental interactions at the highest conceivable energies~\cite{LiteBIRD:2022cnt,CMB-S4:2022ght}.

\section{Baryogenesis}

One of the most fundamental questions in particle cosmology, and in all of physics, is:  
Why does anything exist at all?
According to our best theories, the Big Bang should have produced equal amounts of matter and antimatter. Yet when we look around, we see a Universe dominated by matter, from the atoms in our bodies to the stars and galaxies across the cosmos. Antimatter, by contrast, is nearly absent, existing only in trace amounts in high-energy processes and experimental laboratories.
This striking imbalance is not just a curiosity; it is a profound clue that new BSM physics must exist. The mechanism responsible for creating this matter-antimatter asymmetry is known as baryogenesis.

Baryogenesis refers to any physical process that, starting from a symmetric state with equal baryons and antibaryons, can dynamically generate a tiny excess of baryons - the particles that make up protons and neutrons - during the early stages of the Universe. This excess, though minuscule (about one part in a billion), is what remains after matter and antimatter annihilated, and it accounts for all the ordinary matter we see today.
This section explores the essential ingredients of baryogenesis, how they fit into the thermal history of the Universe, and the most widely studied theoretical frameworks. It also highlights how ongoing experimental efforts, in collider physics, neutrino physics, and cosmology, are actively probing the viability of these ideas.

\subsection{The Sakharov conditions}

Any successful theory of baryogenesis must satisfy a set of requirements first formulated by Andrei Sakharov in 1967~\cite{Sakharov:1967dj}. These so-called Sakharov conditions are:
\begin{enumerate}
\item \textbf{Baryon number violation:} There must exist physical processes that can change the net baryon number. Otherwise, starting from a baryon-symmetric Universe, no asymmetry can ever arise.
\item \textbf{C and CP violation:} The laws of physics must distinguish between particles and antiparticles. Violation of charge conjugation (C) and the combined charge-parity (CP) symmetry is necessary to produce more baryons than antibaryons.
\item \textbf{Departure from thermal equilibrium:} In equilibrium, particle reactions proceed at the same rate as their reverse, preventing the buildup of an asymmetry. A temporary departure from equilibrium is needed to ``freeze in'' the asymmetry.
\end{enumerate}

Remarkably, all three Sakharov conditions are, in principle, satisfied within the SM: 
baryon number is violated non-perturbatively via electroweak sphalerons; CP violation arises from the CKM matrix in the quark sector; and the early Universe provides out-of-equilibrium conditions during phase transitions. 
However, this turns out to be insufficient. The CP violation in the SM is too small~\cite{Gavela:1993ts}, and the electroweak phase transition is a smooth crossover, lacking the strong first-order dynamics required to efficiently generate an asymmetry. 

Observationally, the baryon asymmetry is encoded in the baryon-to-photon ratio:
\[
\eta_B \equiv \frac{n_B - n_{\bar{B}}}{n_\gamma} \approx 6 \times 10^{-10}
\]
This tiny number has been measured independently via BBN and the anisotropies in the CMB, with both methods yielding consistent results. This confirms that the asymmetry was established well before the formation of the first atoms.
The failure of the SM to generate the observed asymmetry from first principles strongly motivates  BSM physics. Extensions often introduce new CP-violating sources, additional scalar fields, or modified dynamics of the electroweak phase transition. These are frequently connected to other open questions in particle physics, such as neutrino masses, Dark Matter, inflation, and the strong CP problem.

\subsection{Popular baryogenesis frameworks}

Several distinct frameworks for baryogenesis have been proposed. Each one satisfies the Sakharov conditions in different ways and operates at different energy scales. Among the most studied ones are:

\subparagraph{Electroweak baryogenesis}
Electroweak baryogenesis operates at the electroweak scale, around $100$ GeV, and relies on a first-order electroweak phase transition to provide the necessary departure from thermal equilibrium~\cite{Morrissey:2012db}. During such a transition, bubbles of broken electroweak symmetry nucleate and expand. CP-violating interactions in the surrounding plasma lead to the generation of chiral asymmetries in the vicinity of the advancing bubble walls. These asymmetries diffuse into the symmetric phase, where they are processed by sphaleron transitions to produce a net baryon number~\cite{Kuzmin:1985mm}.
For this mechanism to work, two key conditions must be satisfied. First, the electroweak phase transition must be strongly first-order, a condition not met within the SM~\cite{Kajantie:1996mn}. Second, additional sources of CP violation are required, as the CP-violating phase in the CKM matrix is insufficient to generate the observed baryon asymmetry.
Realistic implementations of electroweak baryogenesis therefore demand BSM physics. Common extensions include $N$-Higgs-Doublet Models, the Minimal Supersymmetric SM, and various scalar singlet extensions.
Electroweak baryogenesis is particularly compelling because it is experimentally testable. The required new particles often lie within the energy reach of the Large Hadron Collider (LHC) or future colliders, and the mechanism is intimately connected to Higgs physics and the nature of electroweak symmetry breaking.

\subparagraph{Leptogenesis}
Leptogenesis typically operates at very high energy scales, ranging from approximately $10^9$-$10^{13}$ GeV~\cite{Fukugita:1986hr}. The mechanism proposes that the observed baryon asymmetry of the Universe arises indirectly from an earlier generation of lepton asymmetry. In this framework, heavy right-handed neutrinos, or other particles that violate lepton number, decay out of thermal equilibrium in the early Universe, producing a net lepton number. This lepton asymmetry is then partially converted into a baryon asymmetry through non-perturbative electroweak sphaleron processes. A key motivation for leptogenesis is its natural embedding within the seesaw mechanism, which offers an elegant explanation for the smallness of neutrino masses. Several variations of the basic idea exist, including resonant leptogenesis (which can operate at lower energy scales), soft leptogenesis, and flavoured leptogenesis, which accounts for the dynamics of individual lepton flavours.
Leptogenesis is a compelling scenario because it directly links the origin of the baryon asymmetry to neutrino mass generation, an aspect of physics that already hints towards BSM physics. However, it generally occurs at energy scales too high to be probed directly in current or near-future experiments, and thus relies on indirect signatures accessible through neutrino phenomenology and cosmological observations.

\subparagraph{Affleck--Dine baryogenesis}
Affleck--Dine baryogenesis typically operates at supersymmetric or GUT scales~\cite{Affleck:1984fy}. In supersymmetric theories, scalar fields that carry baryon number can acquire large vacuum expectation values in the early Universe. As the Universe evolves, these scalar fields undergo coherent oscillations and eventually decay, generating a net baryon number. This mechanism has the potential to produce large baryon asymmetries and is particularly sensitive to the detailed structure of the scalar potential, as well as to the specifics of supersymmetry breaking. While Affleck--Dine baryogenesis is less directly connected to observable signatures than some other scenarios, it is often studied in the context of string theory, high-scale supersymmetric cosmology, and early-Universe scalar dynamics.

\subsection{Experimental probes of baryogenesis}

Although baryogenesis occurred in the remote past, its consequences, and some of its underlying ingredients, are accessible to modern experiments. One of the most important probes involves precision studies of CP violation. Experiments searching for electric dipole moments (EDMs), such as the ACME collaboration~\cite{Andreev:2018ayy}, place stringent bounds on new BSM sources of CP violation. Similarly, flavour physics experiments like LHCb and Belle II investigate CP violation in the decays of quarks and leptons, searching for anomalies that might hint at the physics responsible for baryon asymmetry.

Collider experiments also play a central role. By probing high energies and producing rare particles, they can search for scalar fields, supersymmetric partners, or other BSM particles that could have participated in baryogenesis. These searches include extended Higgs sectors, new fermions, or any exotic states that couple to the electroweak plasma in ways relevant to baryon number generation.
Neutrino experiments offer an additional, complementary window. Precision measurements of neutrino oscillation parameters help constrain the mixing angles and mass hierarchies relevant for leptogenesis. Searches for neutrinoless double beta decay probe whether neutrinos are Majorana particles - a key ingredient in many baryogenesis models involving lepton number violation.

Cosmological observations also provide critical information. Measurements of the CMB and BBN determine the present-day baryon asymmetry with high precision, while also constraining the presence of additional relativistic species. 
There is growing interest in exploring baryogenesis scenarios through their possible gravitational wave signals, which may be generated during violent early-Universe dynamics such as bubble collisions or plasma turbulence. These signals, if present, could be detected by future experiments such as LISA or the Einstein Telescope, offering a new observational window onto the physical processes responsible for baryogenesis.

\subsection{Baryogenesis and the bigger picture}
Baryogenesis is more than just an explanation for the origin of matter; it is a bridge between particle physics and cosmology, a problem that cannot be addressed without engaging with both domains. It motivates new theoretical models and constrains their parameter spaces. It demands extensions to the SM, and often intersects with Dark Matter, neutrino physics, and Higgs sector dynamics. It offers rich opportunities for interdisciplinary engagement across collider physics, astroparticle physics, and precision cosmology. For any serious researcher entering the fields of particle physics or cosmology, baryogenesis serves as a prime example of how high-energy physics leaves lasting fingerprints on the largest and oldest structures in the Universe.

\section{Dark Matter}

If we were to rely solely on what we can see - stars, gas clouds, luminous galaxies - we would be misled about the true nature of the Universe. Observations across a wide range of astrophysical systems reveal a consistent and striking message: most of the matter in the Universe is ``invisible''. It neither emits nor absorbs light, does not participate in electromagnetic interactions, and yet dominates the gravitational dynamics of galaxies, clusters, and the large-scale structure of the cosmos.
This unseen component is called Dark Matter.
Understanding Dark Matter is one of the most important challenges in modern physics. It lies at the intersection of cosmology, astrophysics, and particle physics, and has implications ranging from the formation of galaxies to the physics of the early Universe. In particle cosmology, Dark Matter is a central player, a relic of the hot early cosmos whose properties must be explained by BSM physics.

\subsection{Evidence for Dark Matter}

The existence of Dark Matter is supported by a wide range of independent observations, all pointing to the presence of mass that cannot be explained by visible matter alone. One of the earliest and most striking clues comes from galactic rotation curves: stars orbiting the outskirts of galaxies move at unexpectedly high speeds, implying the presence of a massive, invisible halo. Gravitational lensing provides further evidence, as light from background galaxies is bent more strongly than can be accounted for by luminous matter, especially in massive structures like galaxy clusters.

The CMB offers a complementary view, capturing the Universe at just 380,000 years old. Its anisotropies encode the relative contributions of photons, baryons, neutrinos, and Dark Matter, with precision measurements from the Planck satellite indicating that roughly 26\% of the energy budget of the Universe is in the form of Dark Matter~\cite{Ade:2015xua}. Large-scale structure observations also require Dark Matter: the formation of galaxies and cosmic filaments - the so-called cosmic web - is accurately reproduced in simulations only when cold, collisionless Dark Matter is included.
Perhaps the most visually compelling evidence comes from merging galaxy clusters like the Bullet Cluster. There, the hot gas observed in X-rays is spatially offset from the bulk of the gravitational mass, as mapped by gravitational lensing, showing that most of the matter involved does not interact electromagnetically, a defining property of Dark Matter.

\subsection{What could Dark Matter be?}

From the perspective of particle physics, Dark Matter must be non-luminous, stable over cosmological timescales, non-baryonic, and composed of particles that were slow-moving
at the time of structure formation. These criteria rule out all SM particles as viable Dark Matter candidates. A wide range of theoretical candidates have been proposed, spanning many orders of magnitude in mass, interaction strength, and production mechanism.
Among the most studied are Weakly Interacting Massive Particles (WIMPs), with masses in the GeV-TeV range~\cite{Jungman:1995df}. These were once in thermal equilibrium with the SM bath in the early Universe and decoupled via freeze-out, leaving behind a relic abundance consistent with observations, a coincidence often referred to as the WIMP miracle. WIMPs arise naturally in many BSM models, such as supersymmetry, extra dimensions, and extended Higgs sectors.

Axions, originally introduced to solve the strong CP problem in QCD, are another leading candidate. Unlike WIMPs, they are produced non-thermally through field oscillations and behave as cold, ultra-light Dark Matter, potentially forming a Bose--Einstein-like condensate. They are searched for using resonant cavity experiments known as haloscopes.
Sterile neutrinos, which mix weakly with active neutrinos, could also play the role of Dark Matter. Depending on their mass and production mechanism, they may form Warm Dark Matter and produce observable X-ray signals via their decay. Other proposals include ultralight scalar fields with masses around $10^{-22}$ eV, which behave more like waves than particles on galactic scales and may help address small-scale structure issues.

Beyond particle candidates, more exotic ideas involve composite or macroscopic objects, such as Q-balls, dark baryons from hidden sectors, or primordial black holes. While many of these are constrained by existing data, none have been definitively ruled out. The variety of proposals reflects both the richness of theoretical possibilities and the fact that the true nature of Dark Matter remains one of the most compelling open questions in physics.

\subsection{Dark Matter production in the early Universe}

Production mechanisms for Dark Matter broadly fall into two categories: thermal and non-thermal. In the thermal freeze-out scenario, Dark Matter was once in thermal equilibrium with the SM plasma. As the Universe expanded and cooled, the interaction rate of Dark Matter fell below the Hubble expansion rate, leading to a freeze-out where annihilations became inefficient and a relic abundance was left behind. The final abundance in this case is primarily determined by the annihilation cross-section.

Alternatively, Dark Matter particles may have been so weakly coupled to the thermal bath that they never reached equilibrium. In this case, known as freeze-in, Dark Matter is gradually produced through decays or scatterings of other particles. This mechanism typically applies to Feebly Interacting Massive Particles (FIMPs), sterile neutrinos, and other candidates with extremely suppressed couplings.
Non-thermal production also includes mechanisms where Dark Matter arises from coherent oscillations of fields that were initially displaced from the minimum of their potential. This is the case for axions and other light scalars. These processes are not governed by temperature or scattering rates but are sensitive to initial field conditions and the scales of spontaneous symmetry breaking.

A final class of scenarios involves gravitational production, in which Dark Matter is generated purely through its coupling to gravity, independent of any direct interaction with the SM. This can occur in the context of inflationary cosmology, especially for super-heavy Dark Matter candidates, often referred to as WIMPzillas. In these models, the rapid expansion and energy density of the early Universe itself can lead to the production of Dark Matter relics, even in the absence of thermal contact with visible matter.

\subsection{The search for Dark Matter}

There are three main experimental strategies used to search for Dark Matter. The first is direct detection, which aims to observe the recoil of a nucleus or an electron caused by the scattering of a Dark Matter particle as it passes through the detector. This approach is pursued by a range of experiments, including liquid xenon detectors such as XENONnT and LUX-ZEPLIN, cryogenic detectors like SuperCDMS and DAMIC, and electron-recoil experiments such as SENSEI, which are sensitive to ultra-light Dark Matter. These experiments require extremely low backgrounds and very high sensitivity, and although tremendous progress has been made, no confirmed Dark Matter signals have been detected so far~\cite{XENON:2023cxc}.

The second approach is indirect detection, which involves searching for SM particles, such as gamma rays, positrons, or neutrinos, that may result from the annihilation or decay of Dark Matter. Key observational targets include the galactic centre, dwarf spheroidal galaxies, and cosmic ray spectra. Experiments like Fermi-LAT, AMS, CTA, and IceCube are actively involved in this strategy. However, interpreting potential signals is challenging due to complex and uncertain astrophysical backgrounds, which can mimic or obscure Dark Matter signatures~\cite{IceCube:2022der}.

The third strategy consists of collider searches, particularly at the LHC, where Dark Matter particles may be produced in high-energy collisions. Because Dark Matter particles do not interact electromagnetically, they escape detection, leading to events with missing transverse energy. Collider searches have placed increasingly strong constraints on simplified Dark Matter models and effective field theories, providing important guidance for model-building even in the absence of direct discovery.

\subsection{Dark Matter and the broader landscape}

Dark Matter is not just an astrophysical puzzle; it challenges the completeness of the SM and motivates a wide range of theoretical ideas, including supersymmetry, hidden sectors with dark forces, extended Higgs sectors, and portal models linking Dark Matter to known particles.
It also intersects with broader questions: Could Dark Matter interact via a ``dark'' force? Could it be related to the baryon asymmetry, as in asymmetric Dark Matter models? Could it explain unresolved astrophysical anomalies?
Research into Dark Matter is highly interdisciplinary. It connects particle theory, experimental searches, astrophysical observations, and cosmological modelling. It is also increasingly data-driven, with new experiments poised to sharpen constraints.
As both an observational reality and a theoretical challenge, Dark Matter remains one of the most compelling clues to BSM physics, guiding efforts to understand the nature of matter and the evolution of the Universe.

\section{Dark Energy}

The twentieth century brought two great revolutions in cosmology. The first was the discovery that the Universe is expanding. The second, even more surprising, was the discovery that this expansion is accelerating.
In 1998, two independent teams studying distant supernovae observed that galaxies were receding from us faster than expected~\cite{Riess:1998cb,Perlmutter:1998np}. This was not simply a faster expansion; it was a qualitatively different kind of behaviour, one that could not be explained by gravity as described in the theory of general relativity with ordinary matter and radiation.
To account for this cosmic acceleration, physicists introduced a new component to the energy budget of the Universe: Dark Energy. It is invisible, non-clumping, and unlike anything known SM phenomenon. Whatever its nature, Dark Energy dominates the energy density of the Universe today, accounting for approximately 70\% of the total.
Understanding Dark Energy is one of the most profound challenges in modern physics. It raises deep questions about the nature of spacetime, the quantum vacuum, and the ultimate fate of the Universe. 

\subsection{What is Dark Energy?}

Dark Energy is not a single theory or particle. It is a general term for the unknown agent driving the accelerated expansion of the Universe.
At a basic level, general relativity allows for such acceleration, wherein field equations contain a term called the cosmological constant, originally introduced by Einstein in 1917 to achieve a static Universe. When this term is positive, it acts as a form of vacuum energy with repulsive gravitational effects, causing space to expand at an accelerating rate.

For decades, the cosmological constant was considered a mathematical curiosity. But after the 1998 supernova results, it gained renewed attention. The simplest explanation for Dark Energy is that it is just a small, positive cosmological constant - a constant energy density pervading all of space, unchanging in time, and with an unusual property: negative pressure.
In the language of fluid dynamics, this means that Dark Energy exerts a tension rather than a push. Unlike matter or radiation, which dilute as the Universe expands, a cosmological constant retains the same energy density even as space stretches. This leads to an accelerating expansion.

\subsection{Observational evidence}
Multiple independent observations converge on the conclusion that the expansion of the Universe is accelerating and cannot be explained by matter alone. The first direct evidence came from Type Ia supernovae, which act as \textit{standard candles}; their observed dimness at high redshift indicates that the expansion of the Universe has accelerated in recent cosmic history. The CMB provides further support: its angular structure, especially as measured by the Planck satellite, reveals a spatially flat Universe, yet the combined contributions of ordinary and Dark Matter fall short of the critical density. This discrepancy suggests that approximately 70\% of the energy content must come from a component with negative pressure, i.e. Dark Energy. Baryon Acoustic Oscillations (BAO), which imprint a regular scale in the distribution of galaxies, serve as another cosmic ruler, and together with CMB and supernova data, reinforce the case for a dominant Dark Energy component. Finally, observations of the growth of large-scale structure, including galaxy clustering and the formation of cosmic filaments, show a suppressed rate of structure formation, consistent with the repulsive effect of Dark Energy accelerating the expansion and slowing gravitational collapse.

\subsection{The cosmological constant problem and possible solutions}
While the cosmological constant is the simplest explanation for Dark Energy, it comes with an enormous theoretical puzzle~\cite{Weinberg:1988cp}.
In quantum field theory, the vacuum is not an absence of fields but rather their lowest-energy state, characterised by persistent zero-point fluctuations. Each quantum field contributes to the total vacuum energy through these fluctuations, even in the absence of real particles. If we sum over all known fields up to a reasonable cutoff (say, the Planck scale), we obtain a vacuum energy density that is at least 120 orders of magnitude larger than the value inferred from cosmology.
This discrepancy - the largest known in physics - is known as the cosmological constant problem. It is not merely a matter of aesthetics; it suggests that our understanding of quantum fields, gravity, or both is fundamentally incomplete.
The problem becomes even more severe when considering renormalization: the bare cosmological constant in Einstein's equations must be fine-tuned to cancel the quantum contributions to extraordinary precision, without any known symmetry or mechanism to enforce such tuning.

Because the cosmological constant appears so unnatural from a theoretical standpoint, many models have been proposed in which Dark Energy evolves over time. These dynamical Dark Energy models typically involve new scalar fields. A simple example is quintessence, where a slowly rolling scalar field drives the accelerated expansion~\cite{Ratra:1987rm}. Unlike a true cosmological constant, the equation of state in such models changes with time, potentially alleviating fine-tuning issues while introducing new dynamical degrees of freedom.
Other frameworks explore more exotic possibilities. In phantom energy models, the equation of state drops below that of a cosmological constant, with $w < -1$, leading to extreme scenarios such as the ``Big Rip'' in which the Universe could end in a future singularity~\cite{Caldwell:1999ew}. K-essence models extend this idea by generalising the kinetic structure of the scalar field, allowing for richer and more flexible dynamical behaviour~\cite{ArmendarizPicon:2000dh}.
To address why Dark Energy becomes dominant only recently, also known as the so-called coincidence problem, some scalar field models incorporate tracker or attractor solutions. These are designed so that the late-time behaviour of the field is largely insensitive to initial conditions, providing a natural mechanism for the onset of cosmic acceleration in the current epoch.

An alternative to introducing a new energy component is to modify the laws of gravity themselves on cosmological scales. Several theoretical frameworks have been developed in this direction. One prominent class is $f(R)$ gravity, which generalises Einstein equations by allowing the gravitational Lagrangian to include nonlinear functions of the Ricci scalar. Another approach arises from braneworld models, inspired by string theory, in which extra spatial dimensions alter gravitational dynamics at large distances. A third example is massive gravity, where the graviton is endowed with a small mass, modifying the long-range behaviour of gravity. These modified gravity theories aim to account for the observed cosmic acceleration without invoking Dark Energy as a separate component. However, they must also satisfy stringent constraints from solar system tests, gravitational lensing observations, and the growth of cosmic structure~\cite{DeFelice:2010aj,Clifton:2011jh,deRham:2014zqa}.

\subsection{Dark Energy in the broader context of particle physics}
From the perspective of particle physics, Dark Energy remains deeply mysterious. Several theoretical ideas have been proposed to bridge this gap. One possibility involves axion-like fields (very light scalars similar to the QCD axion) which can serve as dynamical Dark Energy candidates. Other approaches link the scale of vacuum energy to supersymmetry breaking, although this connection alone does not resolve the fine-tuning problem. A more speculative line of reasoning arises from the landscape of string theory, where a vast array of possible vacuum states exist, each with a different vacuum energy. Within this multiverse picture, the observed value of Dark Energy may simply reflect an \textit{anthropic} selection effect, only in regions with suitable vacuum energy can structure form and observers emerge. While such ideas remain unconfirmed, they illustrate the breadth of creativity that the Dark Energy problem demands from theoretical physics~\cite{Susskind:2003kw}.

\subsection{Future probes and outlook}
To make progress, precision observations are essential. The next generation of cosmological surveys will aim to map the expansion history and the growth of structure with unprecedented detail, seeking signs of deviation from a cosmological constant. Space-based missions such as Euclid will map galaxy clustering and weak lensing across vast cosmic volumes. The Vera Rubin Observatory will track billions of galaxies to study the impact of Dark Energy on structure formation and test modified gravity. Upcoming CMB experiments like CMB-S4 and LiteBIRD will refine our understanding of early-Universe physics, with potential implications for the Dark Energy sector~\cite{Amendola:2016saw,Abell:2009aa,LiteBIRD:2022cnt,CMB-S4:2022ght}. Meanwhile, gravitational wave astronomy offers a promising new avenue by providing independent distance measurements and testing gravity on cosmological scales. Together, these diverse efforts will probe whether Dark Energy evolves over time, whether gravity behaves as expected on large scales, and whether deeper clues to new physics lie hidden in the accelerating cosmos.

Dark Energy is not merely a cosmological curiosity; it is a profound challenge for fundamental physics. It may point toward a new understanding of the quantum structure of spacetime, signal a breakdown of general relativity at large distances, reflect the legacy of early-Universe phase transitions, or reveal something entirely unexpected. Its resolution may require a synthesis of quantum field theory and gravity, a re-examination of the vacuum, or a revision of our assumptions about symmetry and naturalness. Whatever the ultimate explanation, Dark Energy shapes the fate of the Universe and touches upon some of the deepest questions about the nature of reality itself.

\section{Gravitational waves}

For over a century, gravitational waves were a theoretical consequence of general relativity, elegant and mathematically well-founded, but considered practically undetectable. This changed in 2015 when the LIGO observatory made the first direct detection of gravitational waves originating from a binary black hole merger~\cite{LIGOScientific:2016aoc}. That historic observation marked the birth of gravitational wave astronomy, opening an entirely new observational window onto the Universe.

Gravitational waves are ripples in the fabric of spacetime, produced whenever massive objects accelerate asymmetrically, much like how accelerating electric charges emit electromagnetic waves. In general relativity, they travel at the speed of light and carry energy, momentum, and information about their sources. These waves manifest as two transverse polarisation modes (commonly referred to as ``plus'' and ``cross'' modes) which stretch and squeeze space as they propagate. Unlike electromagnetic waves, gravitational waves interact extremely weakly with matter, allowing them to traverse the Universe with minimal absorption or scattering. This makes them ideal messengers, preserving information about the physical processes that produced them. While they were first studied in connection with astrophysical sources such as binary black holes, neutron star mergers, and supernovae, gravitational waves are also expected to arise from cosmological sources linked to the earliest moments of the Universe.

In the context of particle cosmology, gravitational waves offer a unique and powerful probe of high-energy physics. Unlike photons, which provide a picture of the Universe only from the time of recombination, gravitational waves can carry information from much earlier epochs. Because they decouple from matter almost instantly upon generation, they encode a fossil record of high-energy processes such as cosmological phase transitions, inflationary dynamics, the formation and evolution of topological defects, and possible effects of extra dimensions or modified gravity. As such, they provide an unparalleled opportunity to test theories at energy scales far beyond the reach of current or foreseeable particle colliders, potentially probing physics at the GUT or even Planck scale.

\subsection{Sources of cosmological gravitational waves}
Several mechanisms can generate a stochastic gravitational wave background; a persistent, random signal composed of many unresolved sources.

\subparagraph{Inflationary tensor modes}
Quantum fluctuations of the gravitational field during inflation generate a background of tensor perturbations. These primordial gravitational waves are typically predicted to be nearly scale-invariant and Gaussian.
If detected, such a background would provide direct evidence of inflation and could measure the energy scale at which it occurred. The strength of this signal is characterised by the tensor-to-scalar ratio $r$, which is one of the key observables targeted by CMB experiments.

\subparagraph{First-order phase transitions}
Many BSM scenarios predict first-order phase transitions in the early Universe, in contrast to the smooth crossovers at the electroweak and QCD epochs in the SM. In a first-order phase transition, bubbles of the new vacuum phase nucleate within the old phase and expand. As these bubbles grow and collide, they merge and generate bulk flows and turbulence in the surrounding plasma, sourcing gravitational waves. The resulting signals are typically peaked in both frequency and amplitude, with spectral features that depend on the strength of the transition, the velocity of the bubble walls, and the temperature at which the transition occurs~\cite{Caprini:2015zlo,Hindmarsh:2015qta}. First-order transitions are especially prominent in models with extended Higgs sectors, additional scalar singlet fields, or strongly coupled dark sectors. Importantly, if electroweak baryogenesis occurred through a strong first-order electroweak phase transition, it would have generated a stochastic gravitational wave background, making future gravitational wave detectors a direct probe of baryogenesis and electroweak-scale new physics.

\subparagraph{Topological defects}
In theories with spontaneous symmetry breaking, topological defects such as cosmic strings and domain walls may form.
Cosmic strings, in particular, are one-dimensional objects predicted by a wide range of Grand Unified Theories. Oscillating string loops emit gravitational waves, generating a stochastic background or individual bursts~\cite{Vilenkin:2000jqa,Ringeval:2017eww}.
While current limits rule out very heavy strings, lighter networks remain viable. Pulsar timing arrays (e.g. NANOGrav) and space-based interferometers (e.g. LISA) are sensitive to such signatures.

\subparagraph{Preheating and non-perturbative dynamics}
After inflation ends, the inflaton field decays into SM or hidden sector particles. In some models, this occurs via parametric resonance - a violent, non-linear process known as preheating.
This period can involve large inhomogeneities, turbulence, and highly non-equilibrium dynamics, all of which can source gravitational waves.
Such signals are highly model-dependent but provide unique windows into inflaton couplings, reheating dynamics, and the structure of the inflationary potential~\cite{Easther:2006gt,Dufaux:2007pt}.

\subsection{Experimental probes of gravitational waves}
Detecting gravitational waves from cosmological sources requires sensitivity to a much broader range of frequencies than those associated with astrophysical mergers. 
\begin{itemize}
\item CMB polarisation:
At extremely low frequencies (between $10^{-18}$-$10^{-16}$ Hz), imprints of gravitational waves can appear as B-mode polarisation in the CMB, a signature linked to primordial tensor modes from inflation. Experiments such as BICEP/Keck, LiteBIRD, and CMB-S4 aim to detect these subtle polarisation patterns~\cite{LiteBIRD:2022cnt,CMB-S4:2022ght}.

\item Pulsar Timing Arrays:
Moving to the nano-Hertz range ($10^{-9}$-$10^{-7}$ Hz) pulsar timing arrays such as NANOGrav, the European Pulsar Timing Array (EPTA), and eventually the Square Kilometre Array (SKA) look for correlated delays in pulsar signals caused by passing gravitational waves, offering sensitivity to sources like cosmic strings or supermassive black hole binaries~\cite{NANOGrav:2023gor}.

\item Space-based interferometers:
In the milli-Hertz range ($10^{-4}$-$1$ Hz) space-based interferometers like LISA, DECIGO, and the proposed BBO are optimally placed to probe gravitational waves from first-order phase transitions, inflationary relics, and topological defects~\cite{LISA:2017pwj}.

\item Ground-based interferometers: 
At higher frequencies (from $10$-$10^3$ Hz) ground-based interferometers such as LIGO, Virgo, and KAGRA are sensitive to mergers of stellar-mass black holes and neutron stars, and may also detect stochastic backgrounds from exotic high-energy phenomena~\cite{LIGOScientific:2014pky}.
\end{itemize}

Each frequency band probes a different era and mechanism in the early Universe, making multi-band gravitational wave astronomy a powerful tool for particle cosmology.

\subsection{Theoretical tools and challenges ahead}

Theoretical modelling of cosmological gravitational wave signals draws upon a wide array of techniques. These include quantum field theory in curved spacetime, non-equilibrium dynamics, lattice simulations, hydrodynamic modelling, and effective field theory applied to cosmological perturbations. Despite the richness of the framework, uncertainties remain substantial in certain regimes, particularly those involving turbulent reheating or complex phase transitions, but the field is evolving rapidly. Increasing collaboration between cosmologists, field theorists, and numerical physicists is leading to more robust predictions and deeper theoretical insight.

Gravitational waves have revolutionised our view of the cosmos, not only for astronomers, but for physicists seeking to understand the most fundamental forces and constituents of nature. For particle cosmologists, they provide a new observational channel to test the physics of the early Universe, the structure of phase transitions, and the possible existence of hidden sectors, extra dimensions, or topological relics. As new detectors come online across the full frequency spectrum, gravitational waves may soon allow us to explore epochs and energy scales previously thought inaccessible. They are not merely ripples in spacetime, they are echoes from the very origin of the Universe.

\section{Concluding remarks and final reflections}\label{sec:conclusions}

Particle cosmology sits at the intersection of the smallest and largest scales in nature. It weaves together the physics of fundamental particles with the evolution of the Universe, addressing questions that lie at the heart of both disciplines: What were the initial conditions of the Universe? Why is there more matter than antimatter? What is the nature of the invisible matter that holds galaxies together? What drives the acceleration of cosmic expansion? And how can we access energy scales far beyond the reach of terrestrial experiments?
In this chapter, we have explored the key elements of the field, beginning with the thermal history of the early Universe and progressing through inflation, baryogenesis, Dark Matter, Dark Energy, and gravitational waves. Each of these topics links open questions in fundamental physics to cosmological observables, illustrating how insights from particle theory, experimental searches, and astronomical data converge to inform our understanding of the cosmos.
What unifies these topics is the recognition that the SM, while highly successful, is incomplete. The early Universe, with its extreme conditions, offers a natural arena for testing BSM physics.

Several key themes emerge from this landscape. Progress in particle cosmology relies on a genuinely interdisciplinary dialogue, spanning quantum field theory, general relativity, astrophysics, statistical physics, and data science. Students and researchers entering the field must be willing to cross traditional disciplinary boundaries. No single observation or experiment is sufficient on its own; progress comes through the complementarity of probes. Theoretical efforts in model building must balance simplicity and consistency, navigating between minimal extensions and richly structured frameworks that remain viable under experimental and cosmological constraints.
Looking to the near future, we stand on the verge of major observational advances. Projects like LISA, CMB-S4, Euclid, DARWIN, SKA, and DUNE will allow us to test inflation, explore the properties of Dark Energy, and probe possible relics from early-Universe phase transitions. Computational advances are also transforming the field, with tools such as lattice simulations, non-equilibrium dynamics, and machine learning enabling detailed studies of complex phenomena that were previously out of reach.

Despite this progress, many profound questions remain. We do not yet have a complete theory of quantum gravity. The full structure of the scalar sector remains unknown, as does the possibility of hidden symmetries connecting the visible and dark components of the Universe. The mechanisms underlying inflation, baryogenesis, and Dark Matter production are still open to discovery.
Ultimately, the study of particle cosmology reminds us that the Universe is not just a laboratory; it is a living record of its own fundamental laws. Every relic particle, every cosmological imprint, and every ripple in spacetime carries encoded information from epochs we can never revisit directly, but can learn to read their signatures. This chapter offers only a starting point. The literature is vast, the questions are profound, and the tools are evolving. Yet the guiding principle remains unchanged: to understand the Universe is to understand matter, and to understand matter is to trace its journey through the expanding fabric of spacetime.

For young researchers, this is a time of remarkable opportunity. Particle cosmology is a field where deep theoretical ideas can be tested against real data, and where the interplay between theory and observation drives rapid progress. It encourages broad scientific literacy, creative problem-solving, and a curiosity that spans from the quantum to the cosmic. Just as the field itself draws on insights from many disciplines, understanding the early Universe demands new ways of thinking, and new kinds of thinkers. The complexity of the problems we face cannot be solved in isolation. Real progress depends on collaboration, on the exchange of ideas across backgrounds, cultures, and experiences. A homogeneous scientific community risks converging on shared blind spots; diverse perspectives help us challenge assumptions, ask better questions, and see further.
The myth of the lone genius has long haunted physics, but it does not reflect how discovery truly happens. Breakthroughs in particle cosmology, and in science more broadly, emerge not from solitary insight, but from collective effort. They are born from teams that combine different skills, perspectives, and ways of thinking. To uncover the deepest laws of nature, we must build a research community that reflects the full richness of human experience.
And perhaps you, reading this chapter, will be part of that effort. Whether you bring a new idea, a new question, or a new way of seeing an old problem, your contribution may shape the next leap in our understanding of the Universe. The future of particle cosmology will be written by many hands, including yours.

\begin{ack}[Acknowledgments]%
The author acknowledges financial support from Research Ireland Grant 21/PATH-S/9475 (MOREHIGGS) under the SFI-IRC Pathway Programme.
\end{ack}

\bibliographystyle{elsarticle-num}
\bibliography{reference.bib}

\end{document}